
************************************************************************
\input topp.tex

\PHYSREV
\tolerance 2000
\nopubblock
\titlepage

\title{BURGERS TURBULENCE AND INTERFACE GROWTH. SIMILARITY FUNCTIONAL
SOLUTION OF HOPF EQUATION.
THE CASE OF RANDOM GAUSSIAN FORCING. III}

\author{Sergei E. Esipov}

\address{James Franck Institute and Department of Physics, University of
Chicago, 5640 South Ellis Avenue, Chicago, Illinois 60637, USA}
\andaddress{Department of Physics and Material Research Laboratory,
University of Illinois at Urbana-Champaign,
1110 West Green Street, Urbana, Il. 61801-3090, USA}
\vfil
\eject

\abstract{For the problem of Burgers turbulence with random gaussian forcing a
similarity functional solution of Hopf equation is presented
and compared with scaling arguments and replica Bethe-anzatz treatments.
The corresponding field theory is almost non-anomalous.
In one dimension the local fluctuations develop self-similar time-dependent
behavior, while relative fluctuations within the correlation length form a
steady-state with gaussian distribution. This is the precise meaning of the
so-called fluctuation-dissipation theorem. The one-dimensional
properties are also studied numerically.
It is shown that the fluctuation-dissipation theorem is
invalid above one dimension and higher-order cumulants are non-zero.
In two dimensions the cumulants exhibit logarithmic spatial dependence
which is close to but different from that in
the Edwards-Wilkinson case. No other similarity functional solution is found
which may indicate that the ``strong-coupling'' results are not described
by Burgers equation with gaussian noise.}

\chapter{Introduction}

Burgers model of turbulence [1,2], interface growth [3], ballistic deposition
[3,4], domain walls in the anisotropic random-bond Ising model
[5,6], directed polymers [7] and other phenomena
are described by Burgers equation with external conservative stirring
$$\partial_t {\bf v} = \nu \nabla^2 {\bf v} - {1\over2}
{\bf \nabla}v^2 + {\bf \nabla}\eta,\eqn\burg$$
which with the help of the
velocity potential ${\bf \nabla}h=-{\lambda^{-1}}{\bf v}$
gives an equation
$$\partial_t h = \nu \nabla^2 h + {\lambda\over2}({\bf \nabla}h)^2 +
\eta,\eqn\kpz$$
also known as the Kardar-Parisi-Zhang equation [3].

In our previous papers I, II [8] we considered the problem of random initial
conditions and developed the Liouville-like field theory which enables one to
calculate correlators of the field and study their time relaxation.
It was shown that three different classes of initial conditions decay unlike
each other and universality is limited.
In this paper we address the development of Burgers turbulence
starting with the Burgers field at rest ${\bf v}({\bf x},t=0) = 0$ [or
$h({\bf x},0) = 0$] under the
action of external conservative stirring
${\bf \nabla}\eta$ which is distributed
as a gaussian uncorrelated noise in space and time
$$Q[\eta] \sim \exp\Big[-{1\over{4D}}
\int \eta^2 dxdt\Big]. \eqn\grads $$
Spatial cut off $a$ is required in all dimensions to keep the problem
well-defined.

Noise introduces spatial correlations and forms distribution
of the velocity field with the steady-state features at length scales of the
correlation length $L(t)$. At larger distances velocity remains uncorrelated.
The situation is somewhat analogous to the physics of critical phenomena,
say, $\phi^4$-theory [9], provided that the role of time is played by $T-T_c$.
However, the exponents of the $d=1$ version of Burgers turbulence
are simple
fractions, 1/3 and 2/3 [1-3,6]
(to be properly presented below). This provides some hope that in the
associated field theory the anomaly is limited. Let us remind that in critical
phenomena it was the anomalous behavior
of correlators with merging points, first indicated in [10,11],
that resulted in too general
similarity functionals (also known as ``bootstraps'') for the hierarchy of
correlators.
As a result additional unknown constraints were needed. This method was
subsequently in decline due to the invention of $\epsilon$-expansion and
conformal field theory.
Seemingly non-anomalous exponents of the problem \burg-\grads\ suggest that
the situation here
may be different. Note that non-anomalous does not mean
trivial, they are not necessarily
diffusion-like exponents.

In Section 2 we present the similarity
functional for the Hopf equation associated with \burg-\grads. The
Kardar's replica Bethe-anzatz results [6] are re-derived in Section 3 to help
building the explicit similarity functional
in Section 4. As explained in Sections 3,4 the similarity functional contains
correlators which are not present in the results obtained by using the
ground state wave-function of replicated Hamiltonian. Large time limit and
replica limit do not commute, and that is why replica Bethe-anzatz results
are incomplete, although they are remarkably close to the results of
similarity functional solution.
We then discuss the fluctuation-dissipation theorem [12,13]. Section 5
contains the relevant computer simulations in one-dimensional case.
In Section 6 we apply the same ideas to the two-dimensional case.
Only few cumulants are obtained explicitly but the behavior similar to
Edwards-Wilkinson type is evident. There exist a possibility that above two
dimensions the problem \burg - \grads\
becomes trivial. Numerical ``roughening'' exponents are briefly discussed.

\chapter{The Hopf equation and similarity functional}

We shall work with Eq\kpz, and
the field $h({\bf x})$ is conventionally named as
the interface height profile [3].
Let ${\cal P}[h({\bf x}),t]$ be the probability of finding the
interface profile $h({\bf x})$ at time $t$. It obeys the Fokker-Planck equation
(Hopf equation) [9]
$$\partial_t {\cal P}[h({\bf x}),t] = \int dx
\Big\{D{\delta^2\over{\delta{h({\bf x})}^2}} -
{\delta\over{\delta{h({\bf x})}}}\Big[ \nu \partial_x^2 h({\bf x}) +
{\lambda\over{2}}(
\partial_{\bf x}h({\bf x}))^2\Big]\Big\} {\cal P}[h({\bf x}),t].\eqn\hfp$$
The generation functional,
$${\cal G}[J({\bf x}),t] = \int{\cal D}[h]\ {\cal P}[h({\bf x}),t]
\exp\Big[\int J({\bf x})h({\bf x})dx\Big],
\eqn\gen$$
satisfies the Fourier-transformed equation \hfp
$$\partial_t {\cal G}[J({\bf x}),t] = \int dx
\Big\{D J^2({\bf x}) +
J({\bf x})\Big[ \nu \partial_x^2 {\delta\over{\delta{J({\bf x})}}} +
 {\lambda\over{2}}\Big(
\partial_{\bf x}{\delta\over{\delta{J({\bf x})}}}
\Big)^2\Big]\Big\} {\cal G}[J({\bf x}),t].\eqn\hfp$$
Logarithm of the solution ${\cal G}$ is assumed to be
Taylor-expandable
$$\ln {\cal
G} = F_1(t)\int J({\bf x}) dx + {1\over2} \int J({\bf x}_1)J({\bf x}_2)
F_2({\bf x}_1 - {\bf x}_2;t) dx_1dx_2 + $$
$$\sum_{n=3}^{\infty} {1\over{n!}} \int  J({\bf x}_1) ... J({\bf x}_n)
F_n({\bf x}_1 - {\bf x}_2, ..., {\bf x}_{n-1} - {\bf x}_n; t)
dx_1 ... dx_n,\eqn\sol $$
where we used translational invariance of the problem.
The similarity functional anzatz is the assumption that for any $n$ at
length scales exceeding cut off $a$ the cumulants are given by
$$F_n({\bf x}_1 - {\bf x}_2, ..., {\bf x}_{n-1} - {\bf x}_n) = L^{\alpha_n}(t)
f_n\Big({{\bf x}_1 - {\bf x}_2\over{L(t)}}, ...,{{\bf x}_{n-1}
 - {\bf x}_n\over{L(t)}}\Big),\eqn\simil$$
where all spatial dependences are scaled in terms of the correlation
length
$$L(t) = t^{1/z},\eqn\corr$$
which is assumed to obey the power-law behavior.
Exponents $\alpha_n$, $z$ are to be determined together with the similarity
functions $f_n$.
Plugging expansion \sol\ into the Hopf equation we get the
following hierarchy for $F_n$
$${dF_1\over{dt}} = {\lambda\over{2!}}
\partial_{{\bf x}_1}\partial_{{\bf x}_2}
F_2({\bf x}_1 - {\bf x}_2;t)_{| {\bf x}_1 = {\bf x}_2},\eqn\first$$
$$[\partial_t - \nu(\partial_{x_1}^2 + \partial_{x_2}^2)]
F_2({\bf x}_1 - {\bf x}_2;t) = 2D\delta({\bf x}_1 - {\bf x}_2) $$
$$+
{\lambda\over{3!}} [\partial_{{\bf x}_1}\partial_{{\bf x}_3}
F_3({\bf x}_1 - {\bf x}_2, ..., {\bf x}_2 - {\bf x}_3;t)_{|
{\bf x}_1={\bf x}_3} + ...],\eqn\second$$
$$[\partial_t - \nu(\partial_{x_1}^2 + \partial_{x_2}^2 + \partial_{x_3}^2)]
F_3({\bf x}_1 - {\bf x}_2, ..., {\bf x}_2 - {\bf x}_3;t) = $$
$$ +
{\lambda\over{4!}} [\partial_{{\bf x}_1}\partial_{{\bf x}_4}
F_4({\bf x}_1 - {\bf x}_2, ..., {\bf x}_3 - {\bf x}_4;t)_{|
{\bf x}_1={\bf x}_4} + ...] +$$
$$ \lambda{b_{22}}
\{\partial_{{\bf x}_1}F_2({\bf x}_3 - {\bf x}_1;t)\partial_{{\bf x}_2}
F_2({\bf x}_3 - {\bf x}_2;t) + ... \},
\eqn\third$$
and so on, $b_{nm} = {1\over2}[1/(n-1)!(m-1)!]$.
The structure of higher-order equations can be seen from
\first-\third\. In the equation of order $n$ the notation
 ... in square brackets stands for $n$ possible pairings
of the extra argument ${\bf x}_{n+1}$ with all
${\bf x}_1, ..., {\bf x}_n$. The notation ... in curly brackets
stands for $n!$ possible permutations of the indices {\it and}
for all possible binary products of cumulants which result in the
combined order $n+1$. By combined order we mean that, for example,
the product $O(F_2F_2)$ entering Eq\third\ is of the fourth combined order.
Note that the only inhomogeneity in the hierarchy is the noise-related
$\delta$-function in Eq\second. This is a distinct feature of gaussian noise.
More complex noises will be considered elsewhere.

It is important to avoid immediate substitution
of the similarity anzatz \simil\ into the hierarchy, since (generally
speaking) the nonlinear terms contain pairings of arguments, i.e.
request cumulants at separations of its arguments where this anzatz is invalid.
The problem here is twofold. First, we encounter with the distances of order
$a$
where continuous description is no longer applicable, and objects like
$\delta(0)$ must be taken care of. Second, in the anomalous
theories, pairing of arguments may generate new exponents. While first part of
the problem is resolved by introducing cut-off-related
constants, the second part leads to new similarity forms for nonlinear
terms and the hierarchy becomes under-determined, with many possible
solutions.

The situation simplifies in the case of non-anomalous theories, however there
is
in general little hope that the entire hierarchy can be solved as a whole.
In our case, part of the solution of \first-\third\ has been available,
and we found it possible to satisfy the hierarchy.

\chapter{Replica Bethe-anzatz}

The solution of replica Bethe-anzatz by Kardar [6] will be re-derived
and used for discussing linear and time-independent terms in $F_n$s.
For this purpose we apply the Hopf-Cole transformation [1,8] $h={2\nu\over{
\lambda}}\ln w$ and find that $w$ obeys the equation
$$\partial_t w = \nu \nabla^2 w + {\lambda\over{2\nu}} \eta w,\eqn\cole$$
which is sometimes interpreted as the transfer-matrix equation for the
Boltzmann weight of a directed polymer [13], then time plays the role of
longitudinal direction. Solving Eq\cole, and making the inverse transform
we obtain
$$h({\bf x},t) = {2\nu\over{\lambda}}\ln\Big\{ \int dx_0\int_{{\bf x}(0) = {\bf
x}_0}
^{{\bf x}(t) = {\bf x}} {\cal D}[{\bf x}]
\exp\Big\{-\int_0^t dt'\Big[{1\over{4\nu}} \Big({d{\bf x}\over{dt'}}\Big)^2 +
{\lambda\over{2\nu}}\eta({\bf x}(t'),t')\Big]\Big\}\Big\}.\eqn\eich$$
The beginnings of the trajectories entering
\eich\ are uniformly distributed over entire space to ensure
$h({\bf x},0) = 0$. For convenience we remind the dimensions of the
variables and parameters, $[\nu] = L^2/T$, $[D] = L^{d+2}/T$, $[\lambda]
= L/T$, $[h] = L$, $[\eta] = L/T$, $[w]=1$, where $L$ and $T$ stand for
length and time, respectively.

In order to calculate different
correlators of $h({\bf x},t)$ in the method of replicas the logarithm entering
\eich\ is replaced by $\ln w = {1\over{n}}(w^n -1)_{|_{n\rightarrow{0}}}$,
and the distribution \grads\ is employed to compute averages.
One finds that if $m$ trajectories ${\bf x}(t)$ intersect
at a particular site ${\bf x},t$, the averaging gives
$$\int_{-\infty}^{\infty} {d\eta\sqrt{a^d}\over{\sqrt{4\pi{D}}}} \exp\Big(
{\lambda{m}\eta\theta\over{2\nu}} - {\eta^2a^d\theta\over{4D}}\Big) =
\exp\Big({\lambda^2m^2D\theta\over{4\nu^2a^d}}
\Big),\eqn\aver$$
where $a$ and $\theta$ are the spatial and temporal cut-offs, respectively.
The problem is independent on the temporal cut-off.
Note that the result \aver\ is model-sensitive. Usage of different
noise distributions may result in completely different replica
interactions. This also stems from the fact that the spatial cut-off is
explicitly present in height correlators.
The term $m^2$ is the signature of gaussian noise, it
indicates pair interactions of replicated trajectories.
Kardar suggested to write it as $2[{1\over{2}}m(m-1)+{1\over2}m]$, i.e. to
decompose \aver\ into linear weights and pair weights, the latter being
proportional to the number of pairs ${1\over{2}}m(m-1)$.
Let us calculate the simplest correlator $\langle{h({\bf x},t)}\rangle$
first. It is given by
$$\langle{h({\bf x},t)}\rangle = {2\nu\over{\lambda}}\lim_{n\rightarrow0}
[\langle w^n\rangle -1] =$$
$$  {2\nu\over{\lambda}}
{1\over{n}}
\Big[\prod_{j=1}^{j=n}\int dx_{0,j}
\int_{{\bf x}_j(t) = {\bf x}_{0,j}}^{{\bf x}_j(t)={\bf x}}{\cal D}[{\bf x}_j]
\exp\Big( - \int^t L_n dt'\Big) - 1 \Big]
_{|_{n\rightarrow{0}}},$$
$$L_n = \sum_{j=1}^{j=n}\Big[{1\over{4\nu}}
\Big({dx_j\over{dt'}}\Big)^2 + {\lambda^2D\over{4\nu^2a^d}}\Big] +
{\lambda^2D\over{2\nu^2}} \sum_{i < j} \delta({\bf x}_i-{\bf x}_j),
\eqn\lagr$$
where $L_n$ can be interpreted as the $n$-particle Lagrangian.
Consider the $n$ particle wave-function
$$\Psi({\bf x}_1, ..., {\bf x}_n;t) =
\prod_{j=1}^{j=n} \int dx_{0,j}
\int_{{\bf x}_j(t)={\bf x}_{0,j}}^{{\bf x}_j(t)={\bf x}_j}{\cal D}[{\bf x}_j]
\exp\Big(\int^t L_n dt'\Big),\eqn\wavef$$
which satisfies the Shrodinger equation
$$\partial_t \Psi = \hat{H}_n \Psi + \delta(t),\eqn\shro$$
with the Hamiltonian
$$H_n = \sum_{j=1}^{j=n}\Big(\nu
{\partial^2\over{\partial x_j^2}} + {\lambda^2D\over{4\nu^2a^d}}\Big) +
{\lambda^2D\over{2\nu^2}} \sum_{i < j} \delta({\bf x}_i-{\bf x}_j).
\eqn\kaham$$
The solution of Eq\shro\ can be written as
$$\Psi = \sum_l e^{ - E_{n,l}t} \psi_{n,l},\eqn\eigen$$
where $E_{n,l}$ and $\psi_{n,l}$ are the eigen-values and eigen-functions
of
$$\hat{H}_n\psi_{n,l} = - E_{n,l}\psi_{n,l},\eqn\stat$$
and summation in \eigen\ is performed over all
discrete and continuous states $l$.
In the rest of Section 3 and throughout Section 4 we shall limit ourselves
to the {\it one}-dimensional case.

The ground state of the Hamiltonian \kaham\ was found by Kardar [6], and it
is given by Bethe-anzatz,
$$\psi_{n,0} = \psi_0\exp\Big(-{\lambda^2D\over{2\nu^3}}\sum_{i<j}|x_i -
x_j|\Big),\eqn\bethe$$
with the energy
$$E_{n,0} = - {\lambda^2D\over{4\nu^2a}} n -
{\lambda^4D^2\over{12\nu^5}}n(n^2-1).
\eqn\energr$$
Eqs\bethe, \energr\ can be verified by inspection. Naive generalizations
of this anzatz to, say, two dimensions with $K_0$ modified Bessel
functions instead of exponentials do not work. The wave-function
$\psi_{n,0}$ does not satisfy the initial condition $\psi=1$, so that excited
states have to be invoked as discussed by Bouchaud and Orland [6].
These authors list two major problems with the replica Bethe-anzatz solution
of our problem: (i) moments of $w$ diverge too quickly while moments of $h$ are
well defined, and (ii) excited states grow in time slower than the ground
state,
while
continuous states decay exponentially; this encourages to take the
limit $t\rightarrow\infty$ first and $n \rightarrow 0$ second. We shall see
below that these problems [most probably (ii)] lead to losing terms.
Given that the low excited states are included calculations
become quite difficult and additional model assumptions have been used
by Bouchaud and Orland [6] to obtain the exponents of the
correlation length. It is shown below how to get them directly from
\bethe, \energr\ without employing excited states.

The ground state is the part of the solution and one may be interested in
the part of the final average provided by this state.
We then set the amplitude $\psi_0 = 1$ and find $\Psi(x, ...,x;t) =
\exp(-E_nt)$, so that
$$\langle{h(x,t)}\rangle_0 = {2\nu\over{\lambda}} \lim_{n\rightarrow 0}
[\langle w^n \rangle - 1] = - {2\nu\over{\lambda}} \lim_{n\rightarrow 0}
{E_{n,0}t\over{n}} = \Big({\lambda{D}\over{2\nu{a}}} -
{\lambda^3D^2\over{6\nu^4}}\Big) t .\eqn\hav$$
Similar calculations yield
$$\langle{h(x_1,t)h(x_2,t)}\rangle_0 = \Big({2\nu\over{\lambda}}\Big)^2
\lim_{n,m\rightarrow 0}\langle [w^n(x_1,t)-1][w^m(x_2,t)-1]\rangle =
 - {2D\over{\nu}} |x_1 - x_2|,\eqn\hhav$$
$$\langle{h(x_1,t)h(x_2,t)h(x_3,t)}\rangle_0 = {2\lambda{D}\over{3\nu^2}}
 t,\eqn\hhhav$$
and zero for all higher moments. Index $0$ emphasizes that these results
are obtained with the help of the ground state. Let us discuss them.
First of all, Eq\hav\ predicts
the inversion of growth direction with increase of $\lambda$ at the point
$\lambda = 3\nu^3/aD$, while Eq\kpz\ averaged over space ensures
that $\partial_t\langle{h(x,t)}\rangle$ is non-negative. This phenomena
is cut-off dependent and the second term in \hav\ should be
dismissed in the continuous limit as compared to the first one in
parentheses in the right-hand side of \hav. At the same time this is
a warning that the discrete version of the problem \burg-\grads\ may
have properties different from the continuous version, and influence,
for example, numerical studies at large $\lambda$. The possibility
that the discrete model may have its own behavior was
recognized by Kardar [6].

Eq\hhhav\ indicates that the
celebrated exponent $h \propto t^{1/3}$ is present in the distribution.
One then expects terms $t^{2/3}$ in \hhav\ [14], which are absent. It is
also clear that answers \hav-\hhhav\ are not actually the moments,
they look closer to cumulants, since terms $\langle{h}\rangle^2$
are absent in \hhav, and so on. One may conclude that only the terms
which are linear in time and independent on time are obtained from the
ground state. Barring the incompletness, the replica results happen to be
already sufficient for determining basic exponents if one assumes
global scaling of $h$ fluctuations.
The precise meaning of this global scaling is that
the exponents $\alpha_n$ in \simil\ are proportional to $n$ with
the exception of $\alpha_1$. Then, Eq\hhav\ gives $h \sim L^{1/2}$, Eq\hhhav\
results in $h \sim t^{1/3}$, so that $L \sim t^{2/3}$ is the correlation
length. We shall see in the next Section that expressions for cumulants
of $h$ are very close to \hav - \hhhav. Despite obvious success of replica
method, and help that we obtained from it in the next Section,
we feel that at present stage of understanding it remains
uncontrolled for our problem.

The non-trivial contribution to the third moment \hhhav\
means that the distribution of $h$ at short length scales is {\it not}
gaussian. At first glance this contradicts to the statement that the so-called
fluctuation-dissipation theorem (FDT) applies to Burgers equation within the
correlation length [12,13]. The reader may be interested in explicit checking
of FDT. It can be done with the help of Ref.13, where authors
used scaling arguments and argue that the distribution
$${\cal P}_0[h] \propto \exp\Big[-{\nu\over{2D}}\int (\nabla{h})^2 dx\Big]
\eqn\fdt$$
(in our notations) holds at distances $x \ll L(t)$ in one dimension. It is a
straightforward
task to plug \fdt\ into Hopf equation
\hfp, assuming that the left-hand side vanishes
$\partial_t {\cal P} = 0$ for such a steady-state. Diffusive terms
cancel
$$\Big[D{\delta\over{\delta{h({\bf x})}}} - \nu\nabla^2h({\bf x})\Big]
{\cal P}_0 = 0,\eqn\difcancel$$
and we are left with the integral
$$\int dx {d^2h\over{dx^2}} \Big({dh\over{dx}}\Big)^2.\eqn\error$$
This integral is non-zero above one dimension and, therefore,
the distribution \fdt\ is not a solution. In one dimension
the integral \error\ is evaluated at the
boundaries, and {\it may be} zero, although there is no guarantee so far.
We remind that the boundaries of the integration region
correspond to $x \sim L(t)$
and so far we know nothing about the distribution at such
distances. Making the system periodic does not
help either provided that $L(t)$ does not exceed the system size.

We shall see in the next Section that distribution \fdt\ does
describe {\it relative} fluctuations of the interface in one
dimension, i.e. the fluctuations of the field $h(x,t)-h(x',t)$ for
the separations $|x-x'|$ smaller than the correlation length. As for the
{\it local} fluctuations of $h(x,t)$, they are not in a steady-state regime
and can not be described by \fdt. There is nothing novel in this separation
of local and relative fluctuations - the same scenario takes place for
pure diffusion equation with external noise. However, in the latter case
both are gaussian.

We note in passing that the same distribution \fdt\ may give the
steady-state of relative fluctuations for
the stochastic partial differential equation of the
type \kpz\ with a quite general function of $\nabla{h}$, including the
function $\sqrt{1+(\nabla{h})^2}$ [3].

\chapter{Building the similarity functional}

Let us understand the meaning of Eqs\hav-\hhhav\ in terms of the hierarchy
\first-\third\ and similarity anzatz \simil. With gaussian
forcing the only source term
enters Eq\second, and it is appropriate to consider it first.
If the correlator $\langle{h(x_1,t)h(x_2,t)}\rangle$ at late times or
$|x_1-x_2| \ll
L(t)$ contains the term $\propto|x_1-x_2|$ the diffusional
operator in \second\ gives a $\delta$-function which matches the source term.
The same spatial dependence can be seen from \fdt.
It is conceivable that in the second equation of the hierarchy the singularity
($\delta$-function) is balanced by the diffusion-based
derivative of the second-order
cumulant. Yet,
one could have used $F_3$ to balance the singularity, or both $F_2$ and $F_3$.
However, we traced these extra possibilities further and found contradictions.
Numerical
studies [15] and the studies presented in the subsequent Section
also support brownian-like behavior of the pair correlator
at distances shorter than $L(t)$ with no contribution from higher cumulants.

Time-independent
term in $\langle{h(x_1,t)h(x_2,t)}\rangle$ indicates that $\alpha_2 =1$
and
$$F_2(x_1-x_2;t) = L(t) f_2\Big({x_1-x_2\over{L(t)}}\Big) =
L(t) \Big\{ f_2(0) + {D\over{\nu}} {|x_1-x_2|\over{L(t)}} +
o\Big[{|x_1-x_2|\over{L(t)}}\Big] \Big\},
\eqn\secsim$$
where the expansion is applied at $|x_1-x_2| \ll L(t)$.

Returning to Eq\first\ we find that the term $\partial_{x_1}\partial_{x_2}{F_2}
 _{|_{x_1=x_2}}$ is equal to $(D/\nu)\delta(0)$. This is the familiar object
that must be of order of inverse cut off $a^{-1}$. The numerical
coefficient remains undetermined, e.g. lattice version dependent.
Eq\first\ then gives
$${\alpha_1\over{z}} f_1 L^{\alpha_1-z} = {\lambda{D}\over{2\nu}}\delta(0),
\eqn\firsim$$
i.e.
$$\alpha_1 = z,\qquad f_1 \sim \lambda{D}/2a\nu.\eqn\firres$$
Thus, the hierarchy is broken, and we shall build the rest
of the similarity functional. Let us say a word of caution
from the very beginning
that thus obtained similarity solution is not
necessarily unique, and our non-anomalous solution does not, in principle,
exclude possible anomalous ones. On the other hand, such a freedom
allows to make guesses.

Note, that the usage of the similarity
anzatz in Eq\first\ would result in the term $\partial_{x_1}\partial_{x_2}
{F_2} _{|_{x_1=x_2}} = L^{\alpha_2-2}f_2^{''}(0)$, i.e. provide us with
incorrect scaling, $\alpha_1 - z = -1$.
The discussed difference is induced by the singular
$\delta$-function in Eq\second, and is the only anomaly in the theory.
Given that we balanced $\delta$-function by $F_2$
(and not by $F_3$), we expect no additional singularities. The reason for this
is that $F_2$ will only be differentiated once from now on, it
happens in all the equations of hierarchy starting from the fourth one.
Each time it is balanced by diffusion-based second derivative of the
next order cumulant, which in its turn is ``injected'' further on in the
form of its first derivative, and the singularity
is completely healed. We are then ready to plug the similarity
anzatz into the rest of the hierarchy.

Returning to Eq\second\ at finite distances and using the available
relations among exponents, one finds
$${1\over{z}} L^{1-z}(t)[f_2(y_1-y_2) - (y_1-y_2)f_2^{'}(y_1-y_2)] =$$
$${\lambda
\over{3!}} L^{\alpha_3-2}(t) [\partial_{y_1}\partial_{y_3}  f_3(y_1,y_2,y_3)
_{|_{y_1=y_3}} +
\partial_{y_2}\partial_{y_3} f_3(y_1,y_2,y_3) _{|_{y_2=y_3}}].\eqn\secsim$$
which results in
$$z + \alpha_3 = 3,\eqn\expthird$$
and at short distances
$$\partial_{y_1}\partial_{y_3}  f_3(y_1,y_2,y_3)
_{|_{y_1=y_3}} +
\partial_{y_2}\partial_{y_3}
f_3(y_1,y_2,y_3) _{|_{y_2=y_3}} = {6\over{z\lambda}} f_2(0).\eqn\hint$$
Here we used abbreviated notation for arguments of $f_3$.
Eq\hint\ shows that the third cumulant $f_3$ is probably the
quadratic form of its arguments at short distances,
$$f_3 = f_3(0) +  {2\over{z\lambda}} f_2(0)[(y_1-y_2)(y_1-y_3) +
(y_2-y_1)(y_2-y_3) + (y_3-y_1)(y_3-y_2)],\eqn\threeshort$$
with $f_3(0)$ being the value of the cumulant with all three fields at the
same point. Additional caution is to be exercised here: Eq\threeshort\ is only
one (maybe the simplest one) of many possible
expressions which satisfy Eq\hint.

The next step is to go over to Eq\third, which to the leading order
takes the form
$${1\over{z}}L^{\alpha_3-z}(t) [\alpha_3f_3(y_1,y_2,y_3) - \sum_{i<j}(y_i-y_j)
\partial_{y_i-y_j} f_3(y_1,y_2,y_3) ] =
$$
$$\nu{L^{\alpha_3-2}(t)} (\partial_{y_1}^2+\partial_{y_2}^2+\partial_{y_3}^2)
f_3(y_1,y_2,y_3)) + $$
$${\lambda\over{4!}}
L^{\alpha_4-2}(t) [\partial_{y_1}\partial_{y_4}
f_4(y_1, ...,y_4)_{|_{
y_1=y_4}} + ...] +$$
$$ \lambda{b_{22}}
\{\partial_{y_1}f_2(y_3-y_1;t)\partial_{y_2}
f_2(y_3-y_2;t) + ... \}.
\eqn\thirdsim$$
We now have to choose dominating time dependencies. If one assumes that
$\alpha_3-2$ is to contribute to the main order in Eq\thirdsim, then
the inequality $\alpha_3 \geq 2$ follows from terms proportional to $\lambda$.
Eq\expthird\ gives $z \leq 1$ and we arrive to a contradiction,
$\alpha_3-z > \alpha_3-2$, i.e. $\alpha_3-2$ does {\it not} contribute to the
leading order. One is left with three terms in \thirdsim\ and four possible
choices. Generally speaking,
tracing all choices corresponds to an exponentially
growing pattern of logical steps and may lead nowhere.
We have found in our case that the only acceptable
choice is to admit that all three remaining exponents are equal.
It was also found that the same results for exponents
are obtained if one makes use of the
rule: exponent of the order $n$ has to be determined
from the equation of order $n$ matching all lower-order $\lambda$-dependent
terms. One finds $\alpha_3 - z = 0$ and
$$\alpha_1 = \alpha_3 = z = 3/2.\eqn\resultexp$$
Tracing higher orders one finds
$$ \alpha_n = n/2,\qquad n \geq 2,\eqn\patpok$$
c.f. [16]. The system of equations for the functions $f_n$ remains
unsolvable. However, the behavior of $f_n$ at short distances
can be extracted. Progress in this direction requires steps analogous to
those made in deriving Eq\threeshort. Eq\thirdsim\ becomes
$$f_3(0) =  {\lambda\over{4!}} [\partial_{y_1}\partial_{y_4}
f_4(y_1, ...,y_4) _{|_{y_1=y_4}} + ...] +
{\lambda{D^2}\over{\nu^2}}.\eqn\thirdsim$$
It is unclear from Eq\thirdsim\ if the derivatives of the
fourth order cumulant are non-zero. To find out the answer
we have to go on to the next order
$${1\over{3}}[2f_4(y_1,y_2,y_3,y_4) - \sum_{i<j} (y_i-y_j)
\partial_{y_i-y_j} f_4(y_1,y_2,y_3,y_4) ] =
$$
$$ {\lambda\over{5!}} [\partial_{y_1}\partial_{y_5}
f_5(y_1, ...,y_5)_{|_{
y_1=y_5}} + ...] +$$
$$ \lambda{b_{23}}
\{\partial_{y_1}f_2(y_3-y_1;t)\partial_{y_2}
f_3(y_2,y_3,y_4;t) + ... \}.
\eqn\fourthsim$$
It can be easily checked that $O(f_2f_3)$ term on the right hand side
are non-zero, and therefore $f_4$ or $f_5$ are non-zero.
We again arrive at a logical branching point in building the similarity
functional. If one assumes that $f_4$ is non-zero, then
returning to Eq\thirdsim,
one finds a possible symmetric quadratic form for $f_4$
$$f_4 = f_4(0) + 3\Big[f_3(0) - {\lambda{D^2}\over{\nu^2}}\Big]
\sum_{i<j}^4(y_i-y_j)^2;
\eqn\fourthsimfun$$
possible mixed form for $f_5$ is suggested by Eq\fourthsim
$$f_5 = f_5(0) + 12f_4(0)\sum_{i<j}^5(y_i-y_j)^2 + O(y^3),\eqn\fifthsimfun$$
and so forth.
The latter result implies that high order cumulants do not form
a steady-state distribution at short distances, since in the full cumulant
$F_5$ the quadratic terms \fifthsimfun\ are multiplied by
$L^{\alpha_5-2}(t) = t^{1/3}$, i.e.
grow with time. Although interface width is insensitive to these
corrections, one has {\it no} steady-state
distribution at short distances. The only possibility to avoid this is
to choose
$$f_3(0) = {\lambda{D^2}\over{\nu^2}}.\eqn\kardarwin$$

We performed numerical simulations to extract the behavior of high
order local and non-local cumulants. The results of these simulations are
described in the next Section. We find that {\it relative} fluctuations
are convincingly gaussian at short length scales.
Therefore, the possibility \kardarwin\ is realized, and $f_4$ is zero
{\it together with $f_4(0)$}. It is the fifth order cumulant which
matches $O(f_2f_3)$ term in Eq\fourthsim.
This route leads to the Taylor expansion for high order cumulants of odd
orders
$$f_n = O[y^{(n+1)/2}],\eqn\hiorcum$$
and the non-local part of the full cumulant $F_n$, which is proportional
to $L(t)^{a_n-(n+1)/2} = t^{-1/3}$,
vanishes in the steady state. The local
terms $f_n(0)$ for odd $n=5,7, ...$
are all zero. See next Section for further discussion of numerical results
on high order local cumulants.

At this stage it is appropriate to list the available answer for the
one-dimensional problem. The cumulants in \sol\ are given by
$$F_1 = {\lambda{D}t\over{2a\nu}},$$
$$F_2(x_1-x_2) = t^{2/3}f_2[(x_1-x_2)/t^{2/3}] \rightarrow f_2(0)t^{2/3} +
{D\over{\nu}}|x_1-x_2|,$$
$$F_3(x_1,x_2,x_3) = t f_3[(x_1-x_2)/t^{2/3},...] \rightarrow
{\lambda{D^2}t\over{\nu^2}},$$
$$F_n(x_1,...,x_n) = t^{n/3}f_n[(x_1-x_2)/t^{2/3},...] \rightarrow
0,\eqn\resulting$$
and the correlation length of the problem is $L(t) = t^{2/3}$.
The expressions following the sign $\rightarrow$
are the non-vanishing asymptotic
terms at late times (short distances), $|x_i-x_j| \ll t^{2/3}$ for all $i,j$.
We found only one new term not contained in \hav-\hhhav. Two numerical factors
do not match exactly, although we tried to eliminate possible errors.
The results \resulting\ support the assumption of global scaling used in the
renormalization group method for late times. Namely, if one disregards the
uniform growth described by $F_1$, which can be compensated by adding
a constant to the right-hand side of Eq\kpz, and introduce the
rescalings $x = \sigma{x'}$, $t = \sigma^{3/2}t'$, $h = \sigma^{1/2}h'$
[13] then the results \resulting\ do not change.

\chapter{Numerical simulation of the one-dimensional case}

We performed simple numerical simulations of Eq\kpz\
on Sun Sparc 2 workstation using the discrete scheme [17]
$$h(x,t+\Delta{t})=h(x-1,t)+h(x+1,t)-2h(x,t)+$$
$$\bar{\lambda} \{{1 -
[1+g(x,t)/4]^{-1/2}}\}
+\eta(x,t)\sqrt{\Delta{t}},$$
$$g(x,t) = [h(x-1,t)-h(x,t)]^2 + [h(x+1,t)-h(x,t)]^2.\eqn\numerics$$
All parameters have been scaled into $\bar{\lambda}$.
The stability introduced by this numerical scheme is rather robust
and allowed us to use time steps $\Delta{t}=0.01$ without risk of running
into divergences. In fact, we could have used much larger time steps
but then the fluctuations became strongly noise-dependent.
The size of the $h$-array with periodic boundary conditions
was selected to be 500 lattice units and averaging
over 1000 systems was performed. The numerical noise $\eta(x,t)$ was
selected to be an unsophisticated random number generator of
uniform noise on [-1,1]. The initial condition was selected to be $h(x,0) = 0$.
The typical run up to times t=100 takes a day on the Sun Sparc 2 workstation.

Fig.1 displays time-dependence of moments of the relative interface width
$$W_n(x,t) = \langle[h(x,t)-h(0,t)]^n\rangle^{1/n}\eqn\wid$$
at short length scales $x=10$ lattice units as a function of time.
The curves bend down at $t \sim 30$ and interface width saturates for
$n = 2,4,6,8,10$. Thus, the numerical steady-state is achieved, and
if we scale the widths by the factor $[(2n-1)!!]^{1/n}$ which
represents the number of gaussian pairs, we get that the curves are
nicely superimposed, so that the distribution of relative fluctuations
at short length scales is gaussian.
We believe that the remaining noise is due to insufficient averaging,
in fact, the value $(W_4-W_2)/W_2$ is found to obey the root-mean-square
deviation law.

The time-dependence of local cumulants is shown in Fig.2. The first
cumulant is just the averaged interface height, the second and third
are given by the following formula
$$c_n(t) = \langle[h(x,t)-\langle{h(x,t)}\rangle]^n\rangle,\eqn\cu$$
the expressions for higher ones are a bit more cumbersome,
and we refer the reader to mathematical textbooks.
We see that for $\bar\lambda = 2,10$ and $n = 2,3$
the expected power laws develop after a transient, but are in good agreement
with our expectations. The fourth and fifth {\it numerical} cumulants
are, of course, non-zero, although their behavior is
irregular in time at our level of averaging (1000 systems). Kim, Moore and
Bray [18] used much more extensive averaging and reported non-anomalous
numerical exponents for non-zero high cumulants, see also [19] for a
discussion. As is shown in the previous
Section the non-zero local high cumulants are incompatible with the
steady-state for relative fluctuations. Certainly, this argument does not
necessarily apply to lattice models.
The case of $\bar\lambda=10$ displays a new feature.
Here the lattice effects are more ``organized'' and
rare events associated with sign change of fluctuating cumulants
are intermittent with steady growth of $f_{4,5}$.
We consider this as an indication that
at large enough $\bar\lambda$
the connection with the continuous equation is lost to a greater extent
(rather then the establishing connection). Further numerical analysis
seems to be appropriate to study the evolution of numerical high order
cumulants in different lattice model and the mentioned connection with the
continuous case.

\chapter{The two-dimensional case}

Let us apply the same analytical
method to $d=2$ case. Singularity in Eq\second\
leads to the term
$${D\over{\nu}} \ln{|{\bf x}_1-{\bf x}_2|\over{a}}\eqn\twocut$$
for the second-order
cumulant, $F_2$. It is impossible to represent this expression as a part of
the function of \simil\ type which depends exclusively on
$|{\bf x}_1-{\bf x}_2|/L(t)$ and is multiplied by a power $\alpha_2$
of $L(t)$. This complication is due to the presence of logarithms.
Accounting for logarithms is a delicate issue which sometimes
results in answers correct only with logarithmic precision.
One has to decide first if the correlation length acquires
logarithmic dependences. Our study of this problem with random initial
conditions indicated that the correlation
length does contain logarithms [20],
$$L(t) = t^{1/2}\ln^{1/2}(\nu{t}/a^2).\eqn\corrtwo$$
We have also seen that in one dimension the correlation lengths are the
same for both problems (noise-driven and random initial condition cases).
This is our motivation to use \corrtwo\ for the noise-driven case in
two dimensions as well.

Then only minor modification is needed: the values $f_n(0)$ will be allowed to
have logarithmic time-dependences. The second cumulant acquires the form
$$F_2({\bf x}_1-{\bf x}_2,t) = L^{\alpha_2}(t)
f_2\Big({{\bf x}_1-{\bf x}_2\over{L(t)}}\Big) = $$
$$L^0(t)\Big\{f_2(0) + {D\over{\nu}}
\ln{|{\bf x}_1-{\bf x}_2|\over{L(t)}} +
o\Big[{|{\bf x}_1-{\bf x}_2|\over{L(t)}}\Big]\Big\},\eqn\secsimtwo$$
with $f_2(0) = c_1 + c_2\ln L(t)$, $c_{1,2}$ are cut-off-dependent constants
which make it possible to account for \twocut, and
$\alpha_2=0$.
The first equation of the hierarchy
again reduces to \firsim, \firres\ with $\delta(0) \sim
1/a^2$, $\alpha_1 =z$, and
$$F_1 = {\lambda{D}t\over{2\nu{a}^2}}.\eqn\firsttwo$$
Note, that the same arguments result in linear motion of the averaged interface
height in higher dimensions (with $a^2$ to be replaced by $a^d$).
Returning back to Eq\second\ at finite
separations one obtains an expression
$$c_2{d\ln L(t)\over{dt}} = {\lambda
\over{3!}} L^{\alpha_3-2}(t) [\partial_{{\bf y}_1}\partial_{{\bf y}_3}
f_3({\bf y}_1,{\bf y}_2,{\bf y}_3)
_{|_{{\bf y}_1={\bf y}_3}} +
\partial_{{\bf y}_2}\partial_{{\bf y}_3}
f_3({\bf y}_1,{\bf y}_2,{\bf y}_3) _{|_{{\bf y}_2={\bf y}_3}}].\eqn\secsimtwo$$
Assuming that with logarithmic precision $d[\ln L(t)]/dt = 1/zt =
(1/z)L^{-z}(t)$ one finds
$$z + \alpha_3 = 2.\eqn\expthirdtwo$$
At short distances Eqs\hint, \threeshort\
with two-dimensional arguments are valid.
At the level of Eq\thirdsim\ the above mentioned convention leads to
$\alpha_3 = 0$, and one finds from \expthirdtwo\
$\alpha_1 = z = 2.$
At this stage we have to reconsider the importance of the diffusive terms
in the hierarchy. Tracing
higher orders and using the same rule as in Section 4 we find
$$\alpha_n = 0.\eqn\patpoktwo$$
The obtained $z$-exponent implies $L(t)$ is proportional to  $t^{1/2}$
times arbitrary power of logarithm, which we can not fix by this method.
We think that \corrtwo\ is the correct answer. Returning to the level
of power-laws one may claim that the emerging similarity functional is
essentially what is believed to be the
Edwards-Wilkinson type of behavior [21]. However, the
higher order cumulants differ from
their zero counterparts for the evolution of diffusion equation.
For example, the fourth cumulant is non-zero, it satisfies an equation
$$f_3(0) =  {\lambda\over{4!}} [\partial_{{\bf y}_1}\partial_{{\bf y}_4}
f_4({\bf y}_1, ...,{\bf y}_4) _{|_{{\bf y}_1={\bf y}_4}} + ...] +
{\lambda{D^2}\over{3\nu^2}}\Big\{{({\bf y}_3-{\bf y}_1)({\bf y}_3-{\bf y}_2)
\over{|{\bf y}_3-{\bf y}_1||{\bf y}_3-{\bf y}_2|}} + ...\Big\},
\eqn\thirdsimtwo$$
with two other permutations of indexes in curly brackets.
There is no possibility to absorb the inhomogeneity into $f_3(0)$ as we
managed to do this in one dimension.

The renormalization group
rescaling $x = \sigma{x'}$, $t = \sigma^2t'$, $h = h'$ is
consistent with the available results, although the distribution is
``logarithmic to all orders''. This answer
can not be obtained with the help of the
renormalization group, which is insensitive to logarithmic terms.

The appearance of logarithmic correlator in $d=2$, see Eq\secsimtwo,
is the signature of the transition to the new regime [22]. As we mentioned,
when $z=2$
the diffusional terms can no longer be dismussed; more than that, they
begin to dominate above two dimensions. Thus, the non-linear term is no longer
significant, and results can be inferred from simple diffusion equation.

We now discuss the diverse numerical results [15] on interface
roughness in two dimensions which
reported different exponents of interface width, $W(t) \propto
t^{\beta}$, in the range $0 \leq \beta \leq 0.25$. Numerical schemes
and different solid-on-solid models correspond to different discretizations
of Eq\kpz. ``Strong-coupling'' results are usually obtained with large
non-linearity or strong noise amplitude.  The problem in this regime is
known to be sensitive to lattice formulation or noise distribution
and even has a remarkable phase
transition with the increase of disorder [23]. For instance, the transition
observed numerically by Amar and Family [15] may be the indication that
the behavior of the discretized scheme is no longer relevant to the
continuous Eqs\burg\ - \grads\ with gaussian noise. These results
are perfectly relevant to their discrete model, and, possibly, to the
continious case with more sophisticated noise. At the moment this discussion
expresses just a possible viewpoint.

\chapter{Conclusion}

It seems possible that other field-theoretical models can be treated
by this method. It is not clear in advance that the hierarchy can be analyzed
even if one breaks it by using the correct similarity anzatz introduced
in accordance with symmetry requirements or numerical results. The numerical
results are sometimes difficult to obtain, they are
usually quite under-determined even in two dimensions.
Attempts to minimize the number of
relevant terms at each level of hierarchy lead to exponentially growing
logical schemes with more than one possible solutions.
When more than one similarity solution
exists it is unclear which is actually realized in agreement with
initial and/or boundary conditions. Given that Hopf equation
(Fokker-Planck equation) is linear the stability of the similarity
functional does not depend upon the functional and therefore can not
be used for discrimination between similarity solutions.
By analogy with linear
partial differential equations one may think that properly prepared
sums over possible
(similarity) solutions should meet the initial and boundary conditions
which is, in general, a non-trivial problem.

The results of Section 2-5 indicate that the one-dimensional problem
has the scaling at late times which meets the requirements of the
renormalization group method, notwithstanding the internal
contradictions associated with actual application of this method [24].
In two dimensions, see Section 6,
fields also scale, and from the renormalization-group point of view,
the available solution must be placed into the ``universality
class'' of the pure diffusion equation.

Comparing the results of papers I, II and III we found that the relaxation
of random initial conditions and noise-driven dynamics are
sometimes characterized by equal exponent if distributions of the
random fields which are used for initial conditions and permanent
forcing are the same. For the gaussian distributions in one dimension
the kinetic energy $(\partial{h}/\partial{x})^2$ decays as $t^{-2/3}$,
so that interface height grows as $h \sim t^{1/3}$. The same result
is seen from \resulting\ for the fluctuating part of the noise-driven
problem. The correlation length dependence
$L(t) = t^{2/3}$ is the same for
both problems as we have mentioned. It two dimensions we found it
self-consistent to have $L(t) = t^{1/2}\ln^{1/2}(\nu{t}/a^2)$
in both problems, and
the interface height $h \sim \ln^{1/2}(t)$ is again the
same in both cases [compare \secsimtwo\ and Eq(8.6) of Ref.8, II].
We consider this equivalence as a very strong support of the
results presented above.

Equally consistent with the findings of sensitivity to
random initial conditions presented in I, II are our present
expectations that
different noise distributions may lead to principally
different behavior, including the phase transition [23]. The
presence of sensitivity
was also obtained by Ya. Sinai, Z.-S. She, E. Aurell and U. Frisch
for fractional brownian random initial
conditions [25]. The application of all these ideas to the noise-driven case is
the subject of future work.

\ack{ I am indebted to T.J. Newman for his continuous attention and
interest to this problem and help.
I am thankful to Nigel Goldenfeld for attracting my attention to the
shortcomings of renormalization group and methods of similarity solutions,
to Paul Wiegmann for unexpectedly helpful conversations, to Boris Spivak for
sharing his results prior to publication and many clarifying and
dramatic discussions, and to Mehran Kardar for the references connected
with numerical high order cumulants.
This work was supported in part by the Material Research Laboratories at the
University of Illinois at Urbana-Champaign and at the University of Chicago,
and in part by NSF Grant NSF-DMR-89-20538.}
\chapter{References}

\item{1.} J. M. Burgers, {\em The Non-linear Diffusion Equation }, Reidel,
Dordrecht (1974);
\item{2.} D. Forster, D. R. Nelson, and M. J. Stephen, Phys.\ Rev.\ A, {\bf
16},
732, (1977).
\item{3.} M. Kardar, G. Parisi and Y.-C. Zhang, Phys.\ Rev.\ Lett.,
{\bf 56}, 889 (1986); E. Medina, T. Hwa, M. Kardar, and Y.-C. Zhang, Phys.\
Rev.
\ A, {\bf 39}, 3053, (1989); L.-H. Tang, T. Nattermann, and B.M. Forrest,
Phys.\ Rev.\ Lett., {\bf 65}, 2422, (1990); J. Krug and H. Spohn, in {\it
Solids far from Equilibrium: Growth, Morphology and Defects}, edited by
C. Godreche (Cambridge University Press, Cambridge) 1990.
\item{4.} F. Family and T. Viscek, J. Phys. A {\bf 18}, L75, (1985); F. Meakin,
P. Ramanlal, L.M. Sander, and R.C. Ball, Phys. Rev. A, {\bf 34}, 5091, (1986);
P. Meakin and R. Jullien, J. Phys. (Paris), {\bf 48}, 1651, (1987); R. Jullien
and P. Meakin, Europhys. Lett., {\bf 4}, 1385, (1987).
\item{5.} D. A. Huse and C. L. Henley,  Phys.\ Rev.\ Lett.,
{\bf 54}, 2708 (1986).
\item{6.} M. Kardar, Nucl.\ Phys., {\bf B290}, 582, (1987); J. P. Bouchaud
and H. Orland, J. Stat. Phys., {\bf 61}, 877, (1990).
\item{7.} T. Hwa and D. S. Fisher, cond-mat network preprint.
\item{8.}Paper I: S. E. Esipov, T. J. Newman, Phys.\ Rev.\ E, {\bf 48}, 1046,
(1993); paper II: S. E. Esipov, Phys.\ Rev.\ E, {\bf 49}, 2070, (1994).
\item{9.} J. Zinn-Justin,
{\it Quantum Field Theory and Critical Phenomena}, Claredon Press, Oxford,
1990.
\item{10.} A. M. Polyakov, Zh. Eksp. Teor. Fiz, {\bf 55}, 1026, (1968); [Sov.
Phys.\ JETP, {\bf 28}, 533, (1969).
\item{11.} A. B. Migdal, Zh. Eksp. Teor. Fiz, {\bf 55}, 1964, (1968); [Sov.
Phys.\ JETP, {\bf 28}, 1036, (1969).
\item{12.} U. Deker and F. Haake, Phys.\ Rev.\ A, {\bf 11}, 2043, (1975).
\item{13.} D. A. Huse, C. L. Henley and D. S. Fisher, Phys.\ Rev.\ Lett., {\bf
55}, 2924, (1985).
\item{14.} We remind that numerical study of the so-called interface ``width'',
$W = \langle[h(x_1,t)-h(x_2,t)]^2\rangle = 2\langle{h(x,t)}^2\rangle - 2
\langle{h(x_1,t)h(x_2,t)}\rangle$ have been performed [15]. At large
separations $|x_1 - x_2| \gg L(t)$ one has $W(t) = 2\langle{h(x,t)}^2\rangle -
\langle{h(x,t)}\rangle^2$, i.e. only local contributions matter.
It was observed that $W(t) \propto t^{2/3}$ in one dimension.
\item{15.} D.E. Wolf and J. Kertesz, Europhys. Lett., {\bf 4}, 651, (1987);
J.M. Kim and J.M. Kosterlitz,
Phys. Rev. Lett, {\bf 62}, 2289, (1989);
J.G. Amar and F. Family, Phys.\ Rev.\ Lett., {\bf 41}, 3399, (1990);
B.M. Forrest and L.-H. Tang, Phys. Rev. Lett., {\bf 64}, 1405, (1990);
L.-H. Tang, B.M. Forrest, and D.E. Wolf, Phys. Rev. A, {\bf 45}, 7162, (1992);
T. Ala-Nissila, T. Hjelt, and J.M. Kosterlitz, Europhys. Lett., {\
{\bf 19}, 1, (1992).
\item{16.} A.Z. Patashinsky and V.L. Pokrovskii, Zh. Exp. Teor. Fiz., {\bf 46},
994, (1964); [Sov. Phys. JETP, {\bf 19}, 677, (1964)].
\item{17.} The numerical method and partly the code was developed by
David Y. K. Ko, T.J. Newman and Michael E. Swift, and was kindly provided
by T.J. Newman.
\item{18.} J.M. Kim, M.A. Moore, and A.J. Bray, Phys. Rev. A, {\bf 44}, 2345,
(1991).
\item{19.} E. Medina and M. Kardar, J. of Stat. Phys. {\bf 71}, 967, (1993).
\item{20.} In Chapter VIII of paper II we found the expression for the
edge of the absorbing plane, which is the only possible correlation length-like
scale. Its logarithm was denoted by $\bar{z}$.
\item{21.} S. F. Edwards and D. R. Wilkinson,
Proc. R. Soc. London Ser. A {\bf 381}, 17 (1982).
\item{22.} C.A. Doty and J.M. Kosterlitz, Phys. Rev. Lett., {\bf 69}, 1979,
(1992).
\item{23.} B.Z. Spivak and B.I. Shklovskii in {\it Hopping Transport in Solids
}, volume 28 of Modern Problems in Condensed matter Sciences,
ed. by V.M. Agranovich and A.A. Maradudin, North Holland, 1991.
\item{24.} T.J. Newman, private communications and G.L.Eyink, preprint 1994.
\item{25.} Ya.G. Sinai, Commun. Math. Phys., {\bf 148}, 601, (1992);
Z.-S. She, E. Aurell and U. Frisch, Commun. Math. Phys., {\bf 148}, 623,
(1992).

\chapter{Figure captions}

\item {Fig.1} Time dependence of the interface width moments $W_n$, Eq\wid.
The curves are labeled with the numbers $n$ of corresponding $W_n$.
In case (a) $\bar\lambda = 2$; (b) - $\bar\lambda = 10$.
In case (a) curve for $W_2$ is thick since the rescaled $W_n$ (see text)
collapse on top of it. In the case (b) it happens
at late times, while initially the spatial distribution is clearly
non-gaussian.
This is not a lattice effect.

\item {Fig.2} Time dependence of local cumulants of the interface height.
(a) $\bar\lambda = 2$; (b) - $\bar\lambda = 10$. The straight lines are
best fits, which give the numerical exponents for (a) $\alpha_1/z = 1.01(1)$,
$\alpha_2/z = 0.65(1)$, $\alpha_3/z = 0.97(5)$; and (b)
$\alpha_1/z = 1.02(1)$, $\alpha_2/z = 0.68(1)$, $\alpha_3/z = 1.01(5)$.
Data for the fifth cumulant
(where distinguishable) are shown with slightly elliptical dots.

\end